\documentclass[]{spie}  
\usepackage[]{graphicx}
\usepackage{epstopdf} 
 
\title{Interferometric Observations of Explosive Variables: V838~Mon, Nova Aql 2005 \& RS Oph} 

\author{Benjamin F. Lane\supit{a}, Alon Retter\supit{b}, Joshua A. Eisner\supit{c}, Robert R. Thompson\supit{d},  Matthew W. Muterspaugh\supit{e}, 
\skiplinehalf
\supit{a}MIT Kavli Institute for Astrophysics and Space Research, MIT Department of Physics, 70 Vassar Street, Cambridge, MA 02139\\
\supit{b}Astronomy \& Astrophysics Dept., Penn State University, 525 Davey Lab, University Park, PA 16802-6305\\
\supit{c}U. C. Berkeley, Department of Astronomy, 601 Campbell Hall, Berkeley, CA 94720\\
\supit{d}Michelson Science Center, 100-22 California Institute of Technology, Pasadena, CA 91125\\
\supit{e}Department of Geological \& Planetary Sciences, MS 150-21, California Institute of Technology, Pasadena CA 91125\\
}

\authorinfo{Further author information: (Send correspondence to B.F.L.)\\B.F.L.: E-mail: blane@mit.edu, Telephone: 1 617 253 3429}

\begin{document} 
\maketitle 

\noindent
{\bf Copyright 2006 Society of Photo-Optical Instrumentation Engineers.\\}
This paper will be published in SPIE conference proceedings volume 6268, 
``Advances in Stellar Interferometry.''  and is made available as 
and electronic preprint with permission of SPIE.  One print or electronic 
copy may be made for personal use only.  Systematic or multiple reproduction, 
distribution to multiple locations via electronic or other means, duplication 
of any material in this paper for a fee or for commercial purposes, or 
modification of the content of the paper are prohibited.

\begin{abstract}

During the last two years we have used the Palomar Testbed
Interferometer to observe several explosive variable stars, including
V838 Monocerotis, V1663 Aquilae and recently RS Ophiuchi. We observed
V838 Monocerotis approximately 34 months after its eruption, and were
able to resolve the ejecta. Observations of V1663 Aql were obtained
starting 9 days after peak brightness and continued for 10 days. We
were able to resolve the milliarcsecond-scale emission and follow the
expansion of the nova photosphere. When combined with radial-velocity
information, these observations can be used to infer the distance to
the nova. Finally we have resolved the recurrent nova RS Oph and can
draw some preliminary conclusions regarding the emission morphology.

\end{abstract}



\section{INTRODUCTION}
\label{sect:intro}  

With the improving sensitivity limits of ground-based optical
interferometers it is over time becoming more likely that eruptive
transients such as novae will be bright enough to be observed.
The high angular resolution measurements that can be made
with such systems should allow observers to directly probe some
of the aspects of these explosions. Here we discuss three 
cases where such observations have been made.

\begin{figure}[t!]
\begin{center}
\begin{tabular}{c}
  \includegraphics[height=8cm]{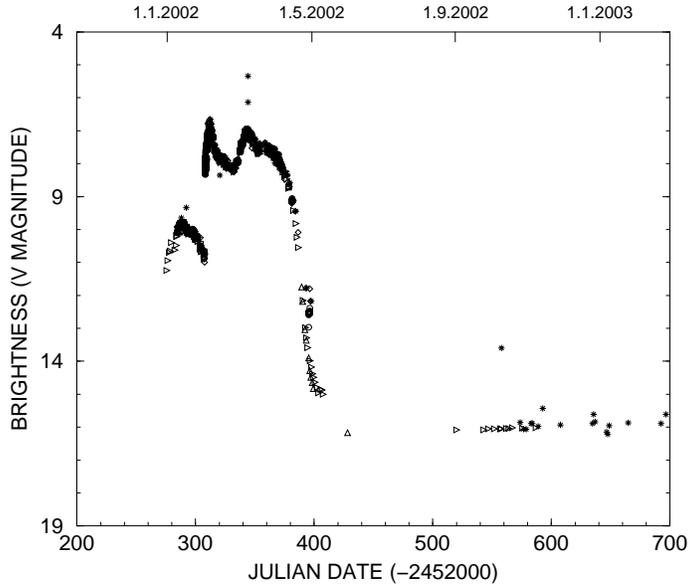}
\end{tabular}
\end{center}
\caption[example] 
{ \label{fig:vband} The V-band light curve of V838 Mon, collected from VSNET circulars.}
\end{figure}

V838 Monocerotis is an explosive variable star that underwent a
nova-like event in early 2002 \cite{b02,mun02}, with a peak magnitude
of $m_V \sim 6.8$ (Fig. \ref{fig:vband}).  However, the eruption was unlike classical novae
in that the effective temperature of the object dropped and the
spectral type evolved into a very late M and L type \cite{evans03}.
The eruption mechanism of V838 Mon is not well understood, but is
probably a new type of explosive variable. There have been many 
models proposed: the merger of a main-sequence binary star \cite{sk03}, 
a He-flash on a post-AGB star\cite{vl04}, or even the accretion of several planets
\cite{rm03}.

Classical novae are energetic stellar explosions that occur in systems
containing a white dwarf (WD) accreting mass from a late-type stellar
companion \cite{hernanz05}. When the amount of accreted material on
the surface of the white dwarf reaches some critical value a
thermonuclear-runaway is ignited, giving rise to the observed nova
outburst in which material enriched in heavy elements is ejected into
the surrounding medium at high velocities.

Nova Aquilae 2005 (ASAS190512+0514.2, V1663 Aql) was discovered on 9
June 2005 by G. Pojmanski \& A. Oksanen\cite{iauc8540}.  At the time
of discovery the magnitude was $m_V$ = 11.05; the source reached $m_V
\sim 10.8$ the following day, and declined in brightness
thereafter. The time to decay 2 magnitudes ($t_2$) was $\sim16$ days,
making V1663 Aql a ``fast'' nova\cite{payne57}. Soon after discovery
M. Dennefeld \& F. Ricquebourg\cite{iauc8544} obtained an optical
spectrum with features indicating a heavily reddened nova. The H-$\alpha$
emission lines exhibited P Cygni line profiles and indicated an
expansion velocity in the range of $700$ km s$^{-1}$ (Dennefeld, personal
communication) to $1000$ km s$^{-1}$ \cite{iauc8640}.

Direct observations of the expansion of the nova shell provide an
opportunity to accurately determine the distance to the nova. Such
observations are usually only possible many years after the outburst,
when the expanding shell can be resolved.  We have used the
Palomar Testbed Interferometer (PTI) to resolve the $2.2 \mu$m
emission from V1663 Aql and measure its apparent angular diameter as a
function of time. We were able to follow the expansion starting $\sim
9$ days after the initial explosion; when combined with radial
velocities derived from spectroscopy we are able to infer a distance
and luminosity of the object. 

Recurrent novae are thought to consist of massive white dwarfs 
orbiting late-type giants; material is pulled from the giant and accreted 
onto the WD. As in the case of classical novae, a thermonuclear runaway 
reaction will on occasion blow off large amounts of matter in a bright,
rapid explosion. There are a small number of systems which have been 
observed through multiple such outbursts, including RS Oph, which 
showed outbursts in 1898, 1933, 1958, 1967, 1985 and 2006. 
It has been argued\cite{hk01} that recurrent novae systems are the progenitors of 
Type Ia supernovae, as the short duration of the explosions indicate 
that the WD should be very close to the Chandrasekhar limit.

\begin{figure}[t!]
\begin{center}
\begin{tabular}{c}
  \includegraphics[height=8cm]{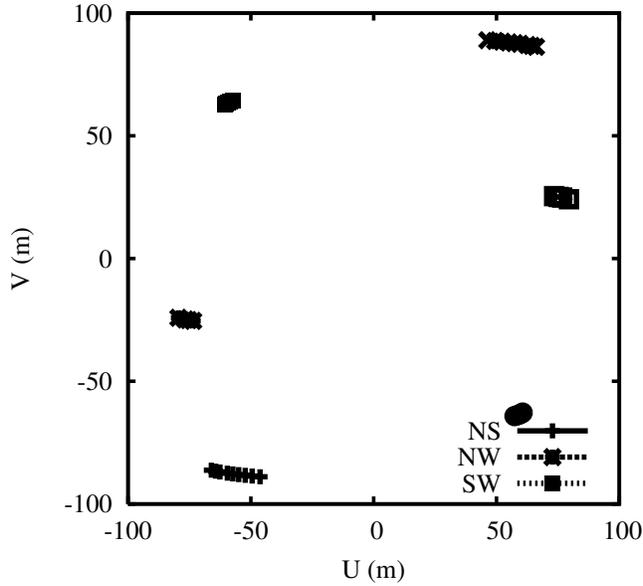}
\end{tabular}
\end{center}
\caption[example] 
{ \label{fig:uv} The $uv$-coverage for one night of observations of V1663 Aql
using PTI.}
\end{figure}

The Palomar Testbed Interferometer (PTI) was built by NASA/JPL as a
testbed for developing ground-based interferometry and is located on
Palomar Mountain near San Diego, CA \cite{colavita99}. It combines
starlight from two out of three available 40-cm apertures and measures
the resulting interference fringes (See Fig. \ref{fig:uv} for representative
 $uv$-plane coverage). The high angular resolution
provided by this long-baseline (85-110 m), near infrared ($2.2 \mu$m)
interferometer is sufficient to resolve emission on the
milli-arcsecond scale.

\section{OBSERVATIONS} 

Each nightly observation with PTI consisted of one or more 130-second
integrations during which the normalized fringe visibility of the
science target was measured. The measured fringe visibilities of the
science target were calibrated by dividing them by the point-source
response of the instrument, determined by interleaving observations of
calibration sources; the calibration sources were chosen to be single
stars, close to the target on the sky and to have angular diameters
less than 2 milli-arcseconds, determined by fitting a black-body to
archival broadband photometry of the sources.  For further details of
the data-reduction process, see Colavita {\it et al.}\cite{colavita99b} and
 Boden {\it et al.}\cite{boden00}.

In addition to measured fringe visibilities, we obtain K-band
photometry using photon count rates from PTI and K-band magnitudes for
the calibrator sources provided by 2MASS \cite{2mass}.  Note that PTI
was not designed with high-precision photometry in mind, and hence the
K-magnitudes should be treated with some caution.

\begin{figure}[t!]
\begin{center}
\begin{tabular}{cc}
  \includegraphics[height=7cm]{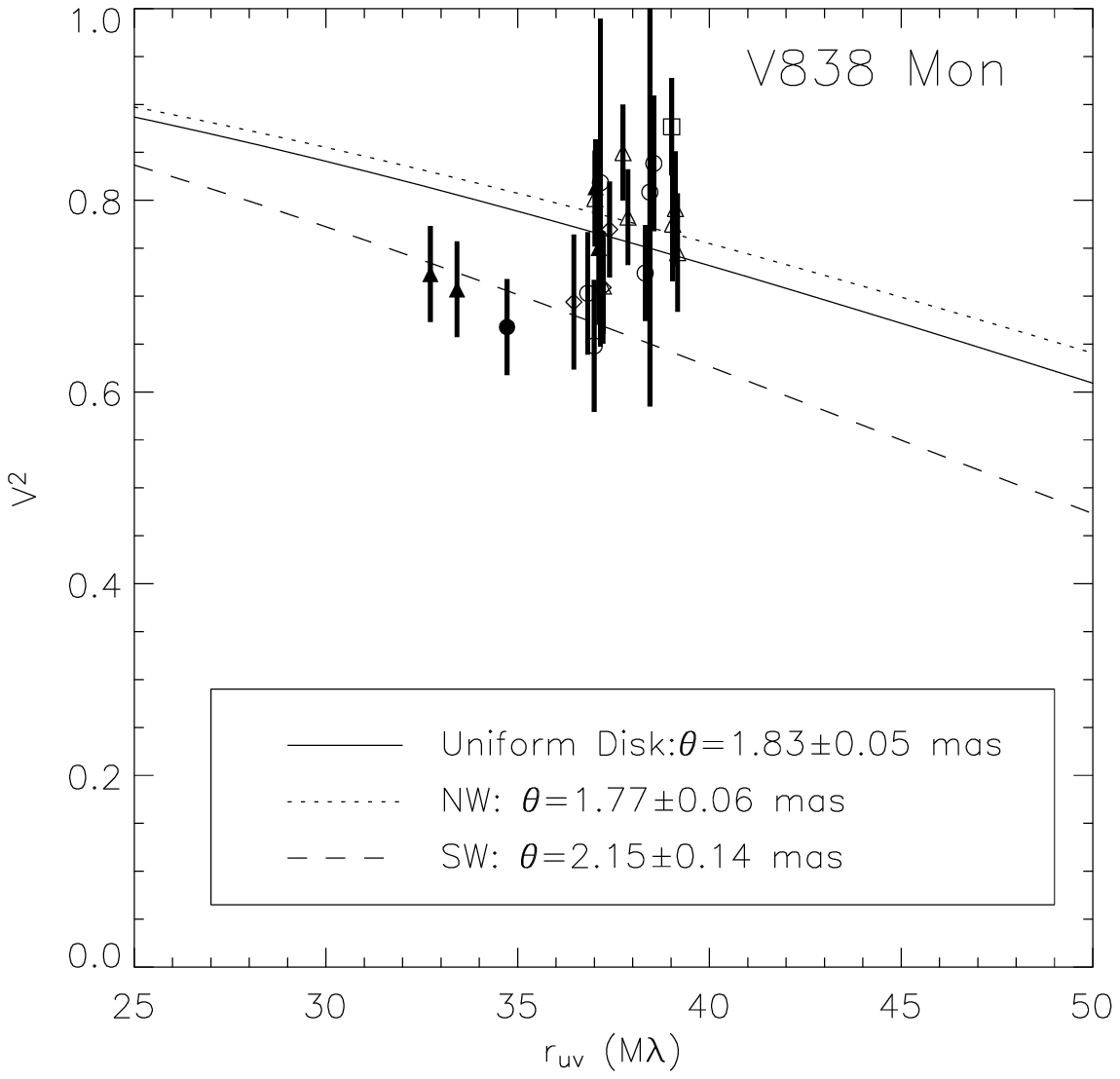}
  \includegraphics[height=7cm]{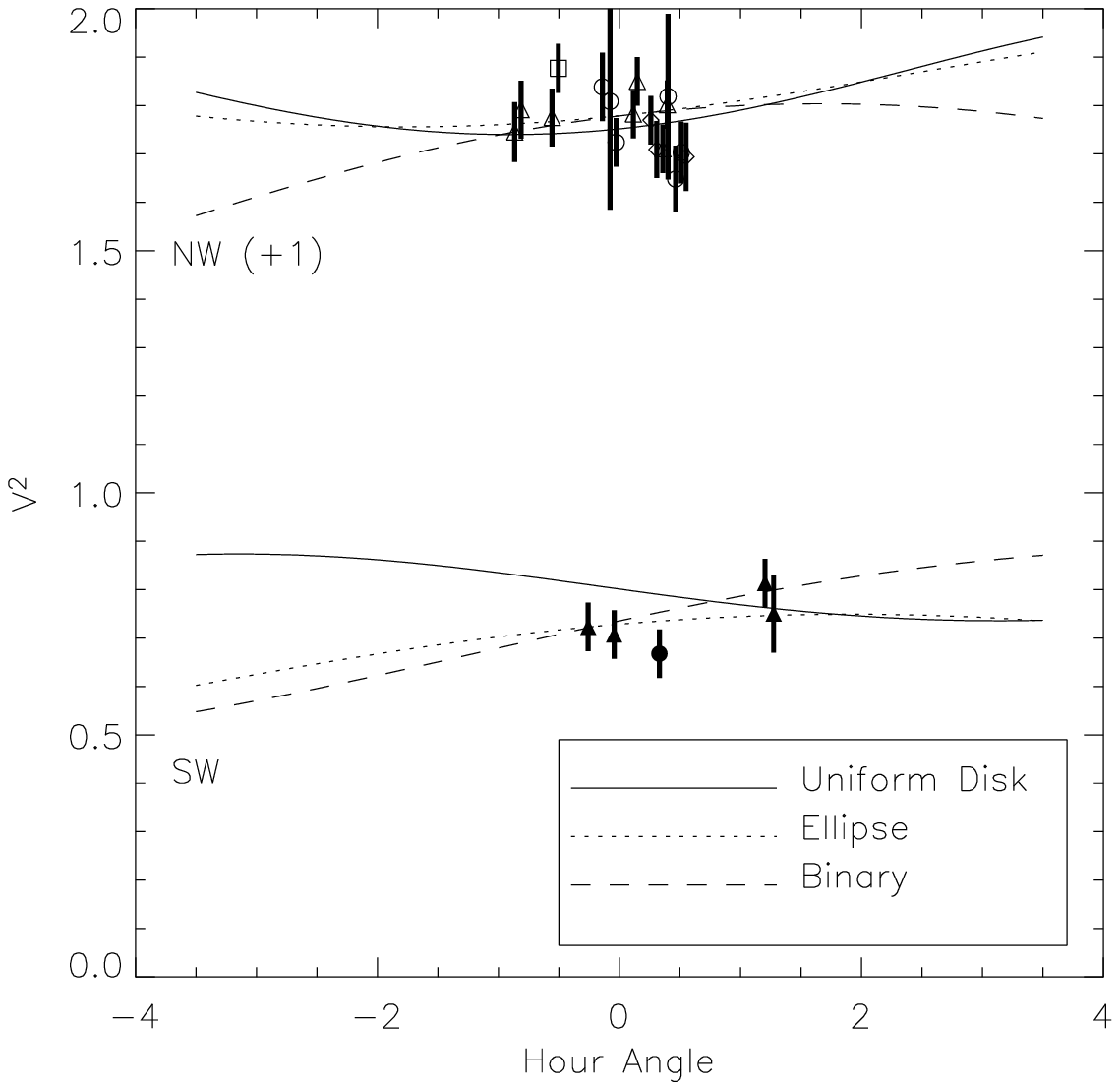}
\end{tabular}
\end{center}
\caption[]{\label{fig:data} Left: The measured fringe visibility
($V^2$) of V838 Mon as a function of projected baseline length
measured in units of the observing wavelength ($2.2 \mu$m). Also shown
are the best-fit Gaussian and uniform disk models, which are
indistinguishable at this projected baseline, but do show that the
emission is resolved and that the data is inconsistent with a
circularly symmetric emission source. Right: The measured fringe contrast
as a function of source hour angle, together with the best-fit
models. A binary or inclined disk model is required to account for the
data from both baselines. Note that the NW-baseline data has been
moved up by 1.0 for clarity. Different symbols are used for different
nights. The figure is reproduced from Lane {\it et al.}\cite{lane05}}
\end{figure}

We observed V838 Mon on 7 nights between 5 November and 
13 December 2004, using PTI in the standard K-band mode, with 
the 85-meter North-West (5 nights) and South-West (2 nights) baselines.

\begin{figure}[t!]
\begin{center}
\begin{tabular}{c}
  \includegraphics[height=8cm]{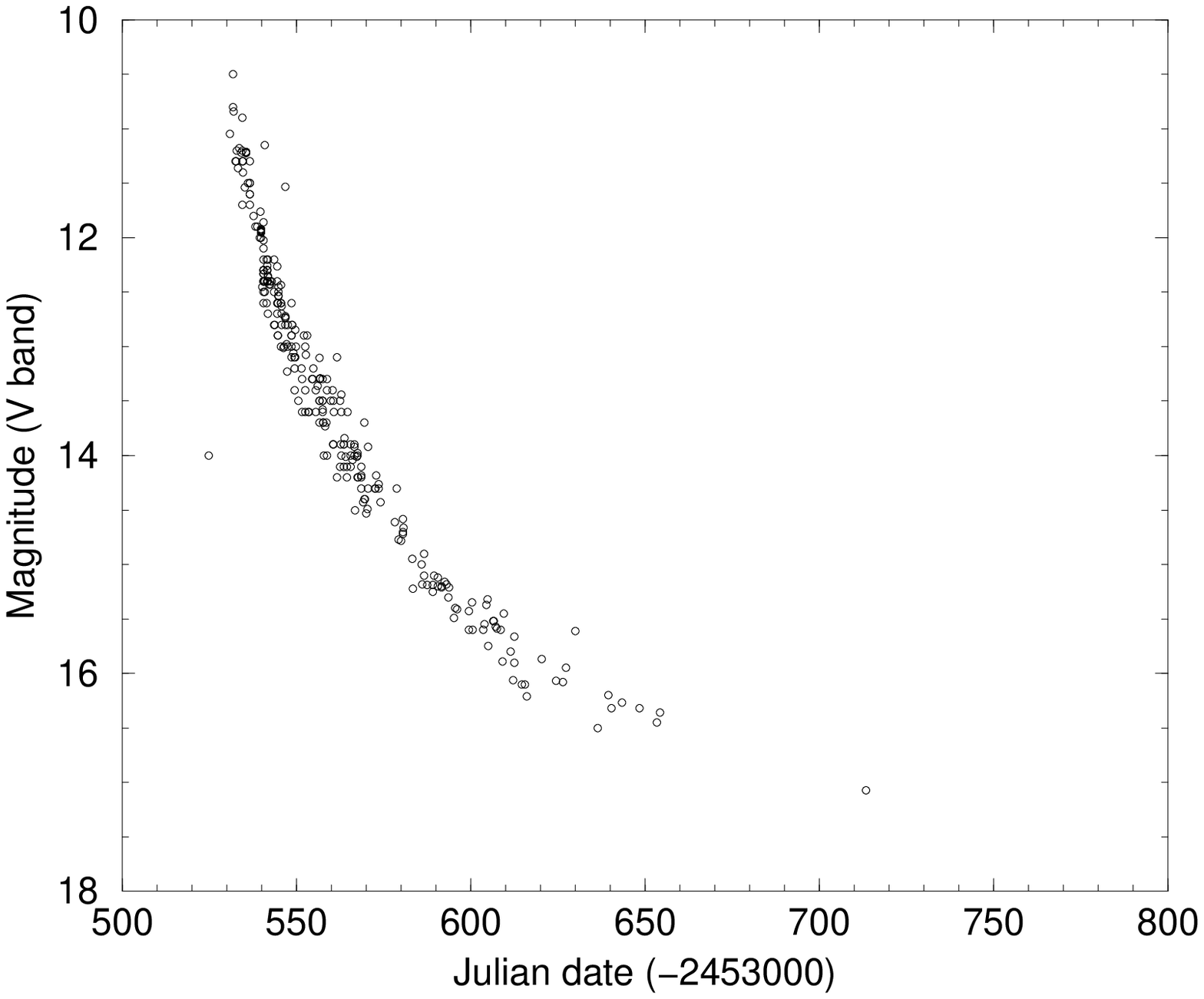}
\end{tabular}
\end{center}
\caption[example] 
{ \label{fig:lcurve} Observed visual light curve 
for V1663 Aql. The observations were collected from 
VSNET publications. The time taken for the V-band lightcurve to 
drop by two magnitudes has been shown to be correlated with 
the maximum source magnitude. For V1663 Aql, the time to drop two 
magnitudes ($t_2$) was $\sim 16$ days.  }
\end{figure}

We observed V1663 Aql on 10 nights between 15 June 2005 and 28 June
2005; on six of those nights we obtained data on two or three
interferometric baselines.  

We observed RS Oph on 2 nights: 24 March and 2 April 2006. Preliminary 
examination indicates that the source is resolved on our 85-meter baseline.

\begin{figure}[t!]
\begin{center}
\begin{tabular}{c}
  \includegraphics[height=9cm]{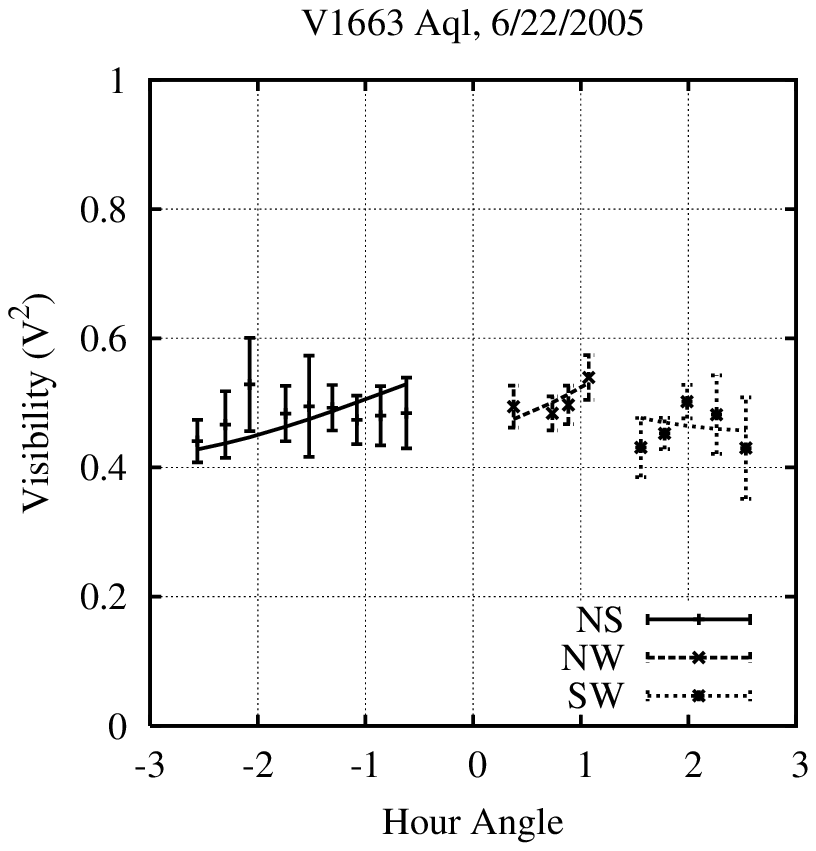}
\end{tabular}
\end{center}
\caption[example] 
{ \label{fig:hour} The measured fringe visibilities of V1663 Aql
on one night, as a function of source hour-angle, 
together with a best-fit inclined disk model (broken lines). }
\end{figure} 

\begin{figure}[t!]
\begin{center}
\begin{tabular}{c}
  \includegraphics[height=9cm]{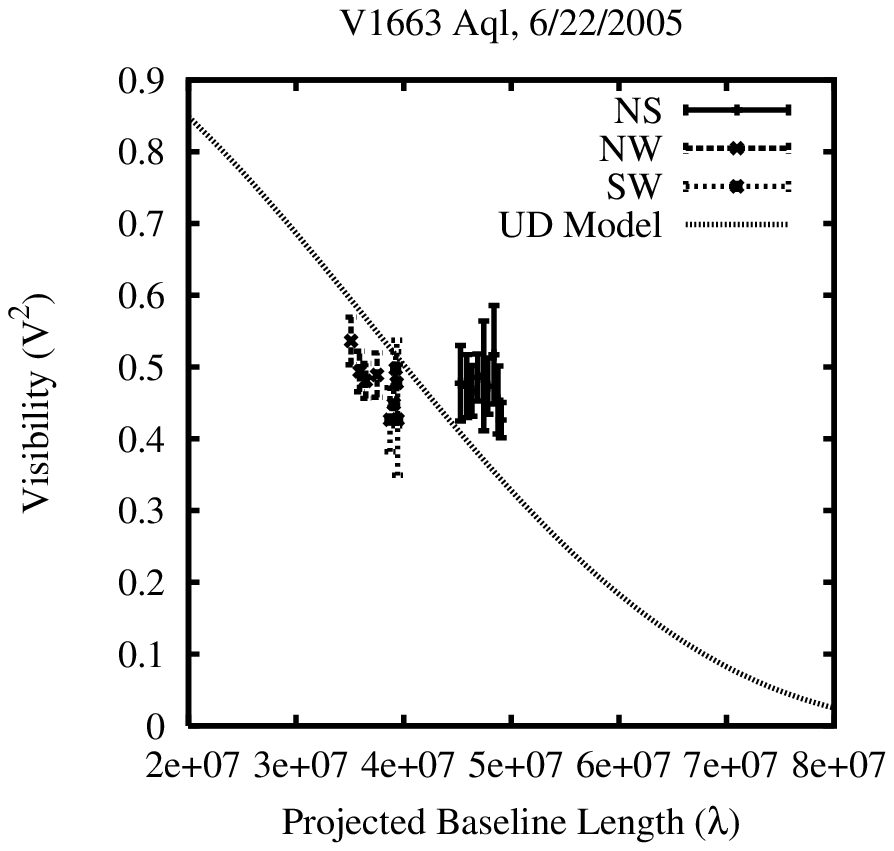}
\end{tabular}
\end{center}
\caption[example] 
{ \label{fig:radplot} The measured fringe visibilities of V1663 Aql 
on one night, as a function of projected baseline length. The
simple uniform-disk model shown does not match the data particularly
well, indicating that the source is more complicated than a simple
uniform disk. }
\end{figure} 

\section{MODELS} 

The theoretical relation between source brightness distribution and
fringe visibility is given by the van Cittert-Zerneke theorem. For a
uniform intensity disk model the normalized fringe visibility
(squared) can be related to the apparent angular diameter using
\begin{equation}
V^2 = \left( \frac{2 \; J_{1}(\pi B \theta_{UD} / \lambda)}{\pi B\theta_{UD} / \lambda} \right)^2
\label{eq:V2_single}
\end{equation}
where $J_{1}$ is the first-order Bessel function, $B$ is the projected
aperture separation and given by $B = \sqrt{u^2 + v^2}$ where $u,v$
are two orthogonal components of the projected baseline, $\theta_{UD}$
is the apparent angular diameter of the star in the uniform-disk
model, and $\lambda$ is the wavelength of the observation.

The availability of data taken with multiple baselines with different
position angles allows us to distinguish between a circularly
symmetric source and elliptical or inclined disk models.  For the
asymmetric cases we fit for three parameters: size ($\theta$),
inclination angle ($\phi$), and position angle ($\psi$). Inclination
is defined such that a face-on disk has $\phi=0$ and $\psi$ is
measured east of north. Following the approach of Eisner {\it et al.}\cite{eis03},
 we include $\phi$ and
$\psi$ in our models of the brightness distribution via a simple
coordinate transformation:
\begin{eqnarray}
u\prime & = & u \sin \psi + v \cos \psi \\
v\prime & = & \cos \phi \left ( v \sin \psi - u \cos \psi \right )
\end{eqnarray}
Substitution of ($u\prime, v\prime$) for ($u, v$) in the expressions
above yields models with inclination effects included.

 We perform least-squares fits of uniform and inclined disk models to
 the measured fringe visibilities in order to derive sizes and inclination angles. 

\section{RESULTS}

\subsection{V838 Mon}

We have published the results of our observations in a recent 
letter\cite{lane05}; we review these results here.
We modeled the $2.2 \mu$m emission from V838 Mon using several
simple emission morphologies, including face-on and inclined disks and
binary models. We find that the best fits to the data are provided by
inclined disk or two-component models. One possibility indicated by
the data is that the observed K-band emission is produced by a very
elongated structure similar to an edge-on disk.  The projected linear
dimensions of such a source are approximately $3.5^{+0.2}_{-1.5}
\times 0.07^{+3.0}_{-0.07}$ milli-arcseconds ($28^{+2}_{-13} \times
0.5^{+24}_{-0.6}$ AU). 

We suggest that the observed emission from V838 Mon was due to ejecta
produced during the eruption. Soon after the peak of the outburst, the
expansion velocity of the ejecta was estimated as 50-350 km/s
\cite{mun02,c03,kip04}. For the time elapsed since outburst of $\sim
10^8$ s, the distance between the ejecta and the star should be in the
range 30--220 AU. This range is consistent with the projected
separations we measure. It is also likely that the ejecta are not
uniformly distributed, and hence the binary morphology preferred by
the data may be ``clumpiness'' in the ejecta.

\subsection{V1663 Aql}

We resolved the emission from V1663 Aql on several nights, and find
that the measured fringe visibilities generally decrease with
time. This implies that the source is expanding.  We also find that
the data on nights with more than one available baseline are not well
matched by a simple uniform disk model, see Fig. \ref{fig:radplot}; in
effect the visibility in the ``North-South'' baseline is too high to
be consistent with the ``North-West'' and ``South-West'' baselines.
There are two morphologies that can fit the observations: the
first is an elliptical disk, with an aspect ratio of $\sim 1.4$:1 (an example of 
such a fit is shown in Fig. \ref{fig:hour}), the second
is a centrally condensed source.  We are currently working to further refine 
models of the emission from this system. 

\begin{figure}[t!]
\begin{center}
\begin{tabular}{c}
  \includegraphics[height=9cm]{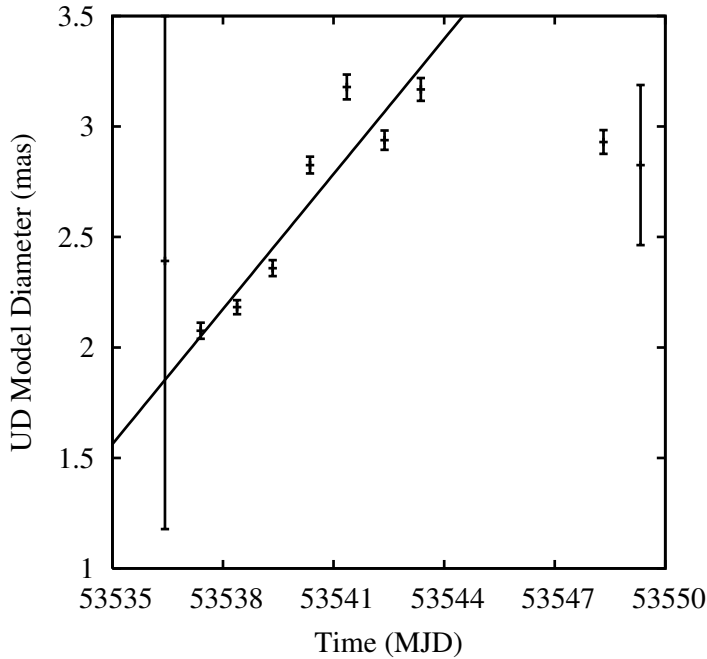}
\end{tabular}
\end{center}
\caption[example] 
{ \label{fig:exp} The best-fit angular size of the emission from V1663 Aql
as a function of time. }
\end{figure}

The measured change in apparent angular size is approximately 0.2 mas/day
(see Fig. \ref{fig:exp});
this can be combined with the observed radial expansion velocity 
inferred from spectroscopic observations to determine the geometric distance
to this nova $d \sim 5.5 \pm 1$ kpc. In addition, by extrapolating the 
linear expansion trend backward in time we can derive the time of the initial 
explosion: MJD $53526.5 \pm 1.3$ d. Finally, we note that the two last data 
points in our series do not follow the simple expansion trend of the previous
points. We suggest that we are seeing the optical depth of the ejecta drop, allowing 
emission from further inside the fireball to escape.

\section{CONCLUSIONS}

In the past two years we have used the Palomar Testbed Interferometer 
to observe three eruptive variable stars: V838 Mon (peculiar variable), 
V1663 Aql (classical nova) and RS Oph (recurrent nova). 
In each case we were able to resolve the emission and measure its 
angular size. In the case of V1663 Aql we 
were able to follow the expansion and derive a geometric distance. 
These observations of V1663 Aql are only the second time  a classical nova has been
resolved by optical interferometry.

\acknowledgments     
 
We wish to acknowledge the extraordinary
observational efforts of K. Rykoski. Observations with PTI are made
possible through the efforts of the PTI Collaboration, which we
gratefully acknowledge. This research has made use of services from
the Michelson Science Center, California Institute of Technology,
http://msc.caltech.edu.  Part of the work described in this paper was
performed at the Jet Propulsion Laboratory under contract with the
National Aeronautics and Space Administration. This research has made
use of the Simbad database, operated at CDS, Strasbourg, France, and
of data products from the Two Micron All Sky Survey, which is a joint
project of the University of Massachusetts and the Infrared Processing
and Analysis Center/California Institute of Technology, funded by the
NASA and the NSF. BFL acknowledges support from a Pappalardo
Fellowship in Physics, JAE is grateful for support form a Miller Fellowship. 
 AR was supported by a research associate fellowship from Penn State University.


\bibliography{report}   
\bibliographystyle{spiebib}   

\end{document}